\documentclass[prc,aps,amsmath,amssymb,nofootinbib,superscriptaddress,12pt]{revtex4-2}
\usepackage{epsfig}
\usepackage[bookmarksnumbered,bookmarksopen,colorlinks,citecolor=blue,linkcolor=blue]{hyperref}
\usepackage{amsmath}
\usepackage{amssymb}
\usepackage{tipa}
\usepackage{fdsymbol}

\begin{document}

    \title{Sub-barrier fusion enhancement due to positive Q-value four-neutron transfer }

\author{Ning Wang}
\email{wangning@gxnu.edu.cn}
\affiliation{Guangxi Key Laboratory of Nuclear Physics and Technology, Guangxi Normal University, Guilin 541004, China }
 
\author{Yi-Jie Duan}
\affiliation{Guangxi Key Laboratory of Nuclear Physics and Technology, Guangxi Normal University, Guilin 541004, China }
 
\author{Hong Yao}
\email{yaohong@gxnu.edu.cn}
\affiliation{Guangxi Key Laboratory of Nuclear Physics and Technology, Guangxi Normal University, Guilin 541004, China }

\author{Hui-Ming Jia}
\affiliation{Department of Nuclear Physics, China Institute of Atomic Energy, Beijing 102413, China }
		
    \begin{abstract}

        The influence of positive $Q$-value four-neutron transfer (PQ4NT) effects on the sub-barrier capture cross sections is systematically investigated using the empirical barrier distribution (EBD2) method. For 13 fusion reactions with $Q_{4n}>0$, sustained neutron-pair transfer is found to reduce barrier heights and enhance capture cross sections at sub-barrier energies. In contrast, reactions such as $^{18}$O+$^{58}$Ni, which have $Q_{2n}>0$ but $Q_{4n}<0$, exhibit no enhancement due to the stalling of subsequent neutron-pair transfer after the initial 2n transfer. By incorporating PQ4NT effects into EBD2 for systems with $Q_{4n}>0$, the average deviation between predicted and experimental capture cross sections (113 datasets) is significantly reduced by $20\%$. Additionally, comparing with the systems induced by $^{48}$Ca ($Q_{4n}<0$), much larger neutron pickup probabilities are observed in the quasi-elastic scattering of $^{40}$Ca-induced reactions ($Q_{4n}>0$) from the time-dependent Hartree-Fock (TDHF) calculations.

    \end{abstract}

     \maketitle

\newpage
  
   \begin{center}
  	\textbf{ I. INTRODUCTION }\\
  \end{center}

The sub-barrier fusion of heavy-ions \cite{Wong73,Bass74,Swiat82,BW91,Gup92,Lei95,Mort99,Jiang24,Denisov,Zhang24}, a process crucial for understanding nuclear structure effects and quantum many-body dynamics in nuclei \cite{Stok78,Das98,Wolski,Jia14}, exhibits complex behavior influenced by couplings to internal degrees of freedom \cite{Hag99,Dasso87}. Among these couplings, the role of neutron transfer channels with positive $Q$-values (PQNT) remains a contentious topic in nuclear physics \cite{Sarg11,Jia12,Zag14,Zhang14,Sarg15,Jia16,Stef17,Wen17,Wangbing,Deb20,Jia25}. While significant sub-barrier fusion enhancement attributed to positive $Q$-value neutron transfer has been proposed in systems like $^{40}$Ca+$^{94,96}$Zr \cite{Stef17}, recent high-precision measurements for $^{18}$O+$^{58}$Ni reveal no such enhancement \cite{Jia25} despite a large positive $Q$-value for two-neutron transfer ($Q_{2n} = +8.20$ MeV). This apparent contradiction underscores the need to identify the precise conditions governing the PQNT effect.

Current understanding suggests that PQNT can enhance sub-barrier fusion through mechanisms such as neutron-rich neck formation \cite{Stel88,ImQMD2014}, which reduces the effective fusion barrier, or by increasing nuclear deformation. It is found that significant deformation increase after neutron transfer correlates with fusion enhancement, while minimal deformation change results in negligible effects \cite{Sarg11,Zhang14}. However, the calculated nuclear deformation is model-dependent (e.g., the FRDM model \cite{Moll95} predicts $\beta_2$ = 0.217 for $^{96}$Zr while WS4 \cite{WS4} suggests near-sphericity), which results in uncertainty in analysis and predictions. Rachkov et al. emphasized that a positive $Q$-value is necessary but insufficient, highlighting the importance of nuclear "rigidity" (resistance to collective excitations) and limiting significant influence primarily to 1n and 2n transfers \cite{Zag14}.

A critical yet unexplored factor may be the continuity of neutron transfer feasibility. Positive $Q$-values for initial transfers (e.g., 2n) may not suffice if subsequent transfers are energetically forbidden. For instance, in $^{40}$Ca+$^{96}$Zr, the $Q$-values $Q_{2n} = +5.53$ MeV and $Q_{4n} = +9.64$ MeV  are favorable for sustained transfer and the enhancement is observed. For $^{18}$O+$^{58}$Ni, $Q_{2n} = +8.20$ MeV but $Q_{4n} = -2.27 $  MeV, which implies neutron transfer stalls after 2n and no enhancement occurs. This suggests that the sign and magnitude of $Q$-values for multi-neutron transfers (particularly 4n) might determine whether neutron neck formation persists, thereby influencing barrier lowering. While the impact of 1n and 2n transfers has been extensively studied, the effect of positive $Q$-value four-neutron (4n) transfer on sub-barrier fusion dynamics remains systematically unaddressed.

This work aims to bridge this gap by investigating the role of 4n transfer $Q$-values in sub-barrier fusion enhancement. We analyze systematics across diverse reaction systems, to explore correlations between the sign of $Q_{4n}$ and fusion enhancement. By integrating the continuity of neutron-pair transfer and nuclear rigidity, we seek to establish predictive criteria for the PQNT effect’s occurrence. Very recently, Wang proposed an analytical formula  (EBD2) \cite{EBD2} with high accuracy for a systematic description of the capture cross sections at near-barrier energies from light to superheavy reaction systems, based on the empirical barrier distribution (EBD) method \cite{SW04,EBD}. In EBD2, the PQNT effect is not yet involved. It is therefore interesting to investigate the influence of the PQNT effect by using the EBD2 predictions as the reference.

 The structure of this paper is as follows: In Sec. II, the framework of the EBD2 formula and the correction terms due to the positive $Q$-value four neutron transfer (PQ4NT) effect will be introduced. In Sec. III, the results from the proposed formula for a series of reaction systems and some discussions are presented.  Finally a summary is given in Sec. IV.

 \begin{center}
	\textbf{ II. EMPIRICAL BARRIER DISTRIBUTION FORMULA }\\
\end{center}
  
 In this work, the capture cross section is written as \cite{EBD2} ,
 \begin{equation}
 	\sigma_{\rm {cap} } (E_{\rm c.m.})=\pi R_B^{2}  \frac{W}{\sqrt{2}E_{\rm c.m.}}
 	[X  {\rm erfc}(-X)+\frac{1}{\sqrt{\pi}}\exp(-X^{2})],
 \end{equation}
where $X =\frac{E_{\rm c.m.}-V_B}{\sqrt{2}W} $.  $E_{\rm c.m.}$ denotes the center-of-mass incident energy. $V_B$ and $W$ denote the centroid and the standard deviation of the Gaussian function, respectively.  $R_B$ denotes the barrier radius. The average barrier height $V_B$ (in MeV) is parameterized as, 
   \begin{eqnarray}
 	V_B=1.051 z +0.000335 z^2 + \Delta B - \Delta_{\rm tr}.
 \end{eqnarray}
$z$ denotes the Coulomb parameter,
  \begin{eqnarray}
 	z=\frac{Z_1 Z_2}{A_1^{1/3}+A_2^{1/3}} F_S
 \end{eqnarray} 
 with $F_S=1-Z_1^{-1/3} Z_2^{-1/3}$. $Z_1$ and $Z_2$  denote the charge number of the projectile nucleus and that of the target, respectively. $A_1$ and $A_2$  denote the corresponding mass number of the reaction partners. The correction term $\Delta B=\sum {\Delta_i I_i^2}$ in Eq.(2) is to consider the competition between shell effect and isospin effect, with the shell gap  $\Delta_i$ \cite{Mo16} and isospin asymmetry $I_i=(N_i-Z_i)/A_i$ of the reaction partners ($i=1$ for projectile and $i=2$ for target). 
In this new version of EBD (v2.2) formula, an additional correction term $\Delta_{\rm tr}=Q_{4n}|I_1-I_2|/2$ is introduced for the reactions with $Q_{4n}>0$ to consider the influence of the PQ4NT effect on the barrier height. Here, $|I_1-I_2|/2$ is to consider the isospin diffusion between the reaction partners.
$Q_{4n}$ denotes the $Q$-value for 4n transfer, which provides an additional energy to transfer neutron pairs and is given by $Q_{4n}=\max(Q_{4n}^{+},Q_{4n}^{-})$ based on the masses of reaction partners. 
\begin{eqnarray}
	Q_{4n}^{+}=M(A_1,Z_1)+M(A_2,Z_2)-M(A_1+4,Z_1)-M(A_2-4,Z_2),
\end{eqnarray}
\begin{eqnarray}
	Q_{4n}^{-}=M(A_1,Z_1)+M(A_2,Z_2)-M(A_1-4,Z_1)-M(A_2+4,Z_2).
\end{eqnarray} 
Considering that $\Delta_{\rm tr}$ should be a small correction to the barrier height, we introduce a truncation, i.e.,  $\Delta_{\rm tr} \leqslant  1$ MeV for  systems with extremely large values of $Q_{4n}$.

The standard deviation of the Gaussian function is parameterized as \cite{EBD2}, 
\begin{eqnarray}
	W=c_0 (1+w_{d}) + c_1 V_B  \sqrt{w_1^2+w_2^2+w_0^2} -\Delta B/N_{\rm CN}^{1/3} +  g \Delta_{\rm tr} ,
\end{eqnarray} 
where $w_{d}=\sum {|\beta_{2i}| A_i^{1/3}}$ and $w_i =  A_i^{1/3} \beta_{2i}^2/(4\pi$), with the mass number $A_1$ and $A_2$ of the reaction partners, and their quadrupole deformation parameters $\beta_2$ taken from the WS4 model \cite{WS4} for prolate nuclei heavier than $^{16}$O. $w_0=(V_B+Q)/c_2$ is introduced to consider the dynamical effects due to the excitation energy at the capture position which is approximately proportional to the excitation energy of a compound nucleus. $Q$
denotes the reaction $Q$-value in fusion process from the ground states of the projectile and target nuclei to the ground state of the compound nucleus. 
$N_{\rm CN}$ denotes the neutron number of the compound nucleus. The values of the model parameters $c_0=0.63$ MeV, $c_1=0.015$ and $c_2=33.0$ MeV are completely the same as those adopted in the previous version \cite{EBD2}. The last term in Eq.(6) is to consider the influence of the PQ4NT effect on the barrier distribution. The factor $g=\left\langle S_{2n} \right\rangle/(V_B+Q)$ is to consider the effect of nuclear “rigidity”.  $\left\langle S_{2n} \right\rangle$ denotes the average two-neutron separation energy of the di-nuclear system. More precisely, $\left\langle S_{2n} \right\rangle=[S_{2n}(A_2,Z_2)+S_{2n}(A_1+2,Z_1)]/2$ if the isospin asymmetry $I_2>I_1$, and otherwise we set $\left\langle S_{2n} \right\rangle=[S_{2n}(A_1,Z_1)+S_{2n}(A_2+2,Z_2)]/2$. The contribution of $\Delta_{\rm tr}$ is relatively larger for the fusion reactions with low excitability (i.e., with smaller excitation energy at capture position), which is consistent with the conclusion in \cite{Zag14}, i.e., the “rigidity” of colliding nuclei with respect to collective excitation is important for sub-barrier fusion enhancement. For fusion reactions with light nuclei such as $^{4}$He, $^{12}$C and $^{16}$O, we note that the influence of nuclear deformation effects on the capture cross sections is weakened due to the large de Broglie wavelength. 
In this work, the value of $W$ is simply written as 
\begin{eqnarray}
	W=c_0 + c_1 V_B
\end{eqnarray} 
for light nuclei ($A\le 16$) induced reactions with $\middlebar{\lambda}_B \ge 0.18$ fm.
Here, $\middlebar{\lambda}_B=\hbar/\sqrt{2\mu V_B}$ denotes the reduced de Broglie wavelength of the colliding nuclei at an incident energy of $E_{\rm c.m.} =V_B$. $\mu$ is the reduced mass of the reaction system.

\begin{table} [!htbp]   	
 	\centering	
	\caption{ Barrier parameters adopted in EBD2.2 for some fusion reactions. Here, $Q_{4n}$ and $Q_{2n}$ denote the corresponding $Q$-value for 4n and 2n transfer, respectively.   }
	\begin{tabular}{cccccc}
		\hline\hline
		
		~~~Reaction~~~  & ~~~$V_B$ (MeV)~~~ &~~~$R_B$ (fm)~~~  &  ~~~W (MeV)~~~  & ~~~$Q_{4n}$ (MeV) ~~~  & ~~~$Q_{2n}$ (MeV) ~~~ \\
		\hline
		$^{18}$O+$^{58}$Ni   &  30.69   &  8.54   &  1.44   &  $-2.27$   &  $+8.20$  \\
		$^{18}$O+$^{116}$Sn  &  49.40   &  9.52   &  1.89   &  $-9.22$   &  $+4.08$  \\
		
		$^{28}$Si+$^{154}$Sm &  99.54   &  10.10  &  3.96   &  $+7.18$   &  $+5.25$  \\
		$^{32}$S +$^{154}$Sm &  112.79  &  10.09  &  4.42   &  $+9.25$   &  $+6.23$  \\
		$^{40}$Ca+$^{154}$Sm &  139.23  &  9.84   &  5.30   &  $+11.22$  &  $+6.01$   \\
		$^{48}$Ca+$^{154}$Sm &  137.12  &  10.45  &  4.27   &  $-5.37$   &  $-2.33$  \\
		
		$^{28}$Si+$^{208}$Pb &  126.16  &  10.42  &  2.88   &  $+5.95$   &  $+4.98$ \\
		$^{32}$S+$^{208}$Pb  &  143.11  &  10.21  &  3.37   &  $+8.01$   &  $+5.95$  \\
		$^{36}$Ar+$^{208}$Pb &  160.15  &  9.83   &  3.85   &  $+8.17$   &  $+6.52$  \\
		$^{40}$Ca+$^{208}$Pb &  176.97  &  9.25   &  4.14   &  $+9.98$   &  $+5.74$  \\
		$^{48}$Ca+$^{208}$Pb &  174.16  &  10.38  &  2.10   &  $-6.60$   &  $-2.60$  \\
		$^{44}$Ti+$^{208}$Pb &  193.84  &  8.46   &  4.83   &  $+14.30$  &  $+8.62$  \\
		
		$^{32}$S+$^{94}$Zr   &  78.42   &  9.50   &  2.54   &  $+6.15$   &  $+5.11$  \\
		$^{32}$S+$^{96}$Zr   &  78.00   &  9.55   &  2.67   &  $+7.66$   &  $+5.74$  \\
		$^{32}$S+$^{104}$Ru  &  85.11   &  9.62   &  3.04   &  $+5.78$   &  $+4.93$  \\
		$^{32}$S+$^{116}$Sn  &  95.80   &  9.71   &  2.88   &  $+1.78$   &  $+2.95$  \\
		$^{32}$S+$^{182}$W   &  131.85  &  10.10  &  4.06   &  $+6.81$   &  $+5.31$  \\
		
		$^{40}$Ca+$^{90}$Zr  &  97.79   &  9.48   &  2.45   &  $-4.18$   &  $-1.45$  \\
		$^{40}$Ca+$^{94}$Zr  &  96.33   &  9.59   &  3.06   &  $+8.13$   &  $+4.89$  \\
		$^{40}$Ca+$^{96}$Zr  &  95.83   &  9.65   &  3.25   &  $+9.64$   &  $+5.53$  \\
		$^{40}$Ca+$^{124}$Sn &  115.68  &  9.88   &  3.85   &  $+9.49$   &  $+5.41$  \\
		$^{40}$Ca+$^{132}$Sn &  114.79  &  10.08  &  4.00   &  $+13.44$  &  $+7.29$  \\
		\hline\hline
	\end{tabular}
\end{table}

 \begin{figure}
 	\setlength{\abovecaptionskip}{ -0.5  cm}
 	\includegraphics[angle=0,width=0.85 \textwidth]{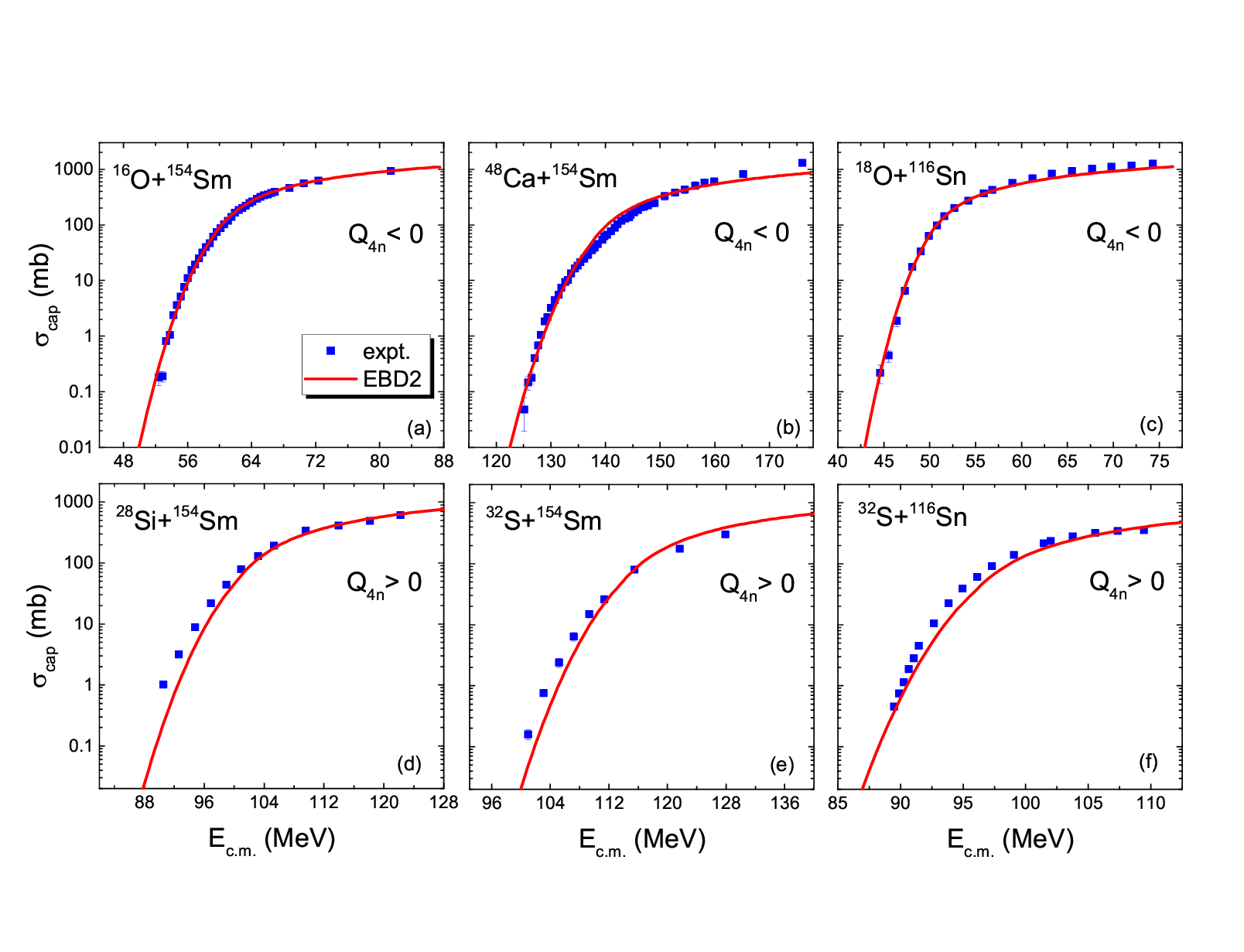}
 	\caption{ Capture excitation functions for $^{16}$O+$^{154}$Sm \cite{Lei95}, $^{48}$Ca+$^{154}$Sm \cite{Stef05}, $^{18}$O+$^{116}$Sn \cite{Deb20}, $^{28}$Si+$^{154}$Sm \cite{Gil90}, $^{32}$S+$^{154}$Sm \cite{Gom94}, $^{32}$S+$^{116}$Sn \cite{Trip01}. The squares and curves denote the experimental data and predictions of EBD2 \cite{EBD2}, respectively.   
 	}
 \end{figure}

\begin{center}
	\textbf{ III. RESULTS AND DISCUSSIONS }\\
\end{center}

\begin{center}
	\textbf{ A. Capture cross sections }\\
\end{center}

We firstly investigate the discrepancies between EBD2 predictions and the measured capture cross sections for $^{154}$Sm and $^{116}$Sn involved fusion reactions. In this work, we mainly focus on medium-mass reaction systems in which the quasi-fission could be negligible and the capture excitation function is therefore approximately equal to the corresponding fusion excitation function. The barrier parameters for some fusion reactions and the corresponding values of $Q_{4n}$ and  $Q_{2n}$ are listed in Table I, in which the barrier heights $V_B$ are obtained by Eq.(2), the values of $W$ are given by Eq.(6), and the barrier radii $R_B$ are taken from Eq.(4) in \cite{EBD2}. For $^{48}$Ca+$^{154}$Sm, $^{48}$Ca+$^{208}$Pb and $^{40}$Ca + $^{90}$Zr, both $Q_{2n}$ and $Q_{4n}$ are negative. For $^{18}$O+$^{58}$Ni and $^{18}$O+$^{116}$Sn, $Q_{2n}>0$ but $Q_{4n}<0$. For other reactions in the table, both $Q_{2n}$ and $Q_{4n}$ are positive.  In Fig. 1, we show the capture excitation functions for $^{16}$O+$^{154}$Sm \cite{Lei95}, $^{48}$Ca+$^{154}$Sm \cite{Stef05}, $^{18}$O+$^{116}$Sn \cite{Deb20}, $^{28}$Si+$^{154}$Sm \cite{Gil90}, $^{32}$S+$^{154}$Sm \cite{Gom94}, $^{32}$S+$^{116}$Sn \cite{Trip01}.  We note that for the three reactions $^{16}$O+$^{154}$Sm, $^{48}$Ca+$^{154}$Sm and $^{18}$O+$^{116}$Sn, the corresponding values of $Q_{4n}$ are negative. For the other three reactions, $Q_{4n}>0$, i.e. the $Q$ values  are favorable for sustained neutron transfer. One can see from the figure that for the reactions with $Q_{4n}<0$, the experimental data can be reproduced remarkably well by the EBD2. Whereas for the reactions with $Q_{4n}>0$, the capture cross sections at sub-barrier energies are slightly under-predicted, which could be due to that the neutron transfer effects are not involved in the EBD2 calculations.

\begin{figure}
	\setlength{\abovecaptionskip}{ -2.5 cm}
	\includegraphics[angle=0,width=0.85 \textwidth]{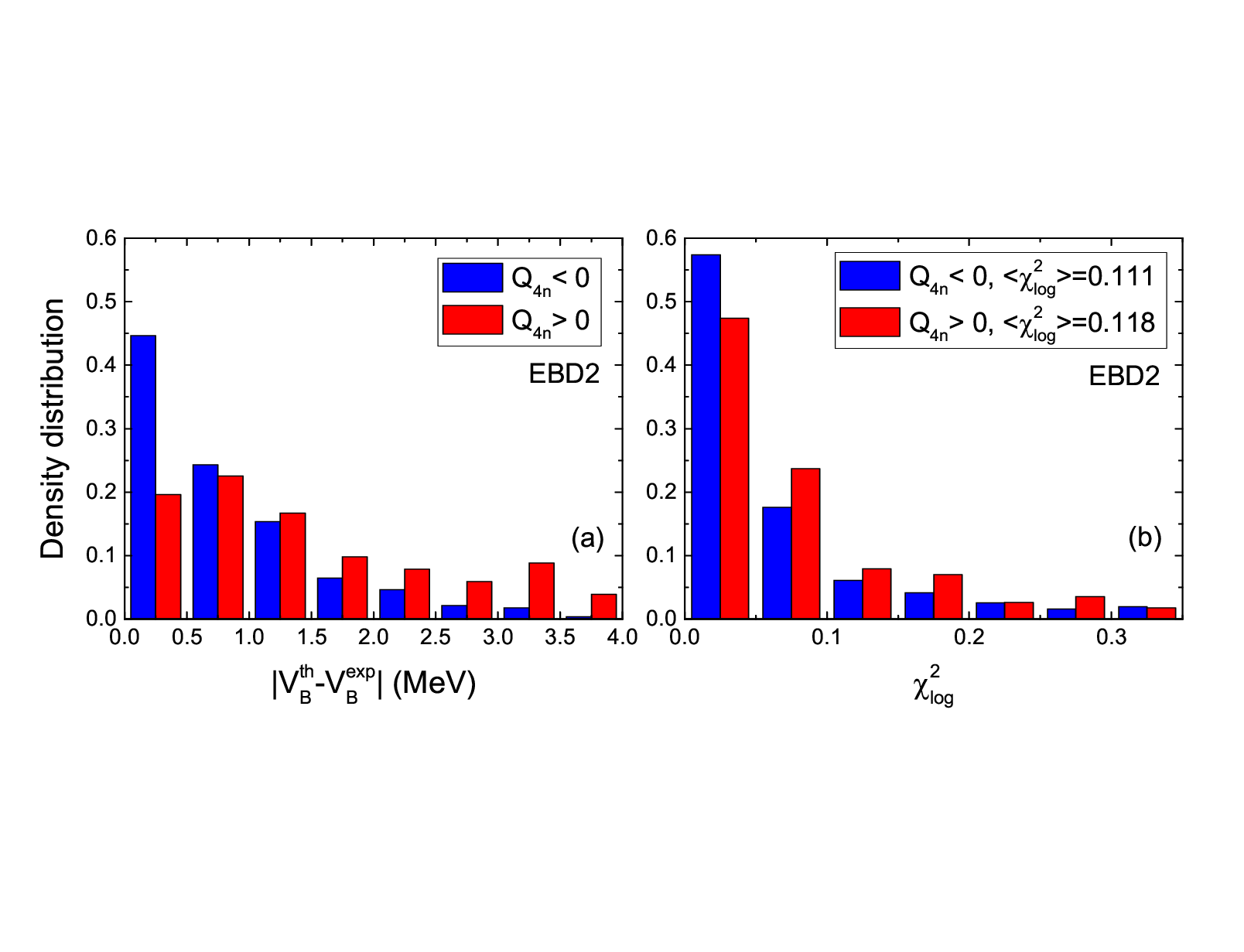}
	\caption{ (a) Distribution of the discrepancies between the calculated barrier heights $V_B^{\rm th}$ with EBD2 and the extracted ones $V_B^{\rm exp}$ \cite{Chen23}. (b) Distribution of the mean-square deviation between the predicted cross sections with EBD2 and the experimental data in logarithmic scale. The blue and the red bars denote the results for the cases with $Q_{4n}<0$ and those with $Q_{4n}>0$, respectively.    }
\end{figure}

For further verifying the correlation between the PQ4NT effect and the sub-barrier enhancement of capture cross sections, we systematically compare the discrepancies between the calculated barrier heights with EBD2 and 382 datasets of extracted values for fusion reactions induced by nuclei with $A\ge 12$ \cite{Chen23}. Simultaneously, the deviations between the predicted capture cross sections with EBD2 and 426 datasets of measured cross sections \cite{EBD} are also explored systematically. In Fig. 2, we show the distributions of the deviations of the barrier heights and the capture cross sections. The blue bars and the red ones denote the results for the cases with $Q_{4n}<0$ and those with $Q_{4n}>0$, respectively. We observe that for reactions with $Q_{4n}>0$, both the discrepancies in barrier heights and the deviations in capture cross sections are systematically larger compared with $Q_{4n}<0$ cases.  While the average mean-square deviation is 0.111 for the reactions with $Q_{4n}<0$, the deviation goes up to 0.118 for the cases with $Q_{4n}>0$, which indicates that the neutron transfer effects are not negligible for the fusion reactions with $Q_{4n}>0$.   

\begin{figure}
	\setlength{\abovecaptionskip}{ 0 cm}
	\includegraphics[angle=0,width=0.8 \textwidth]{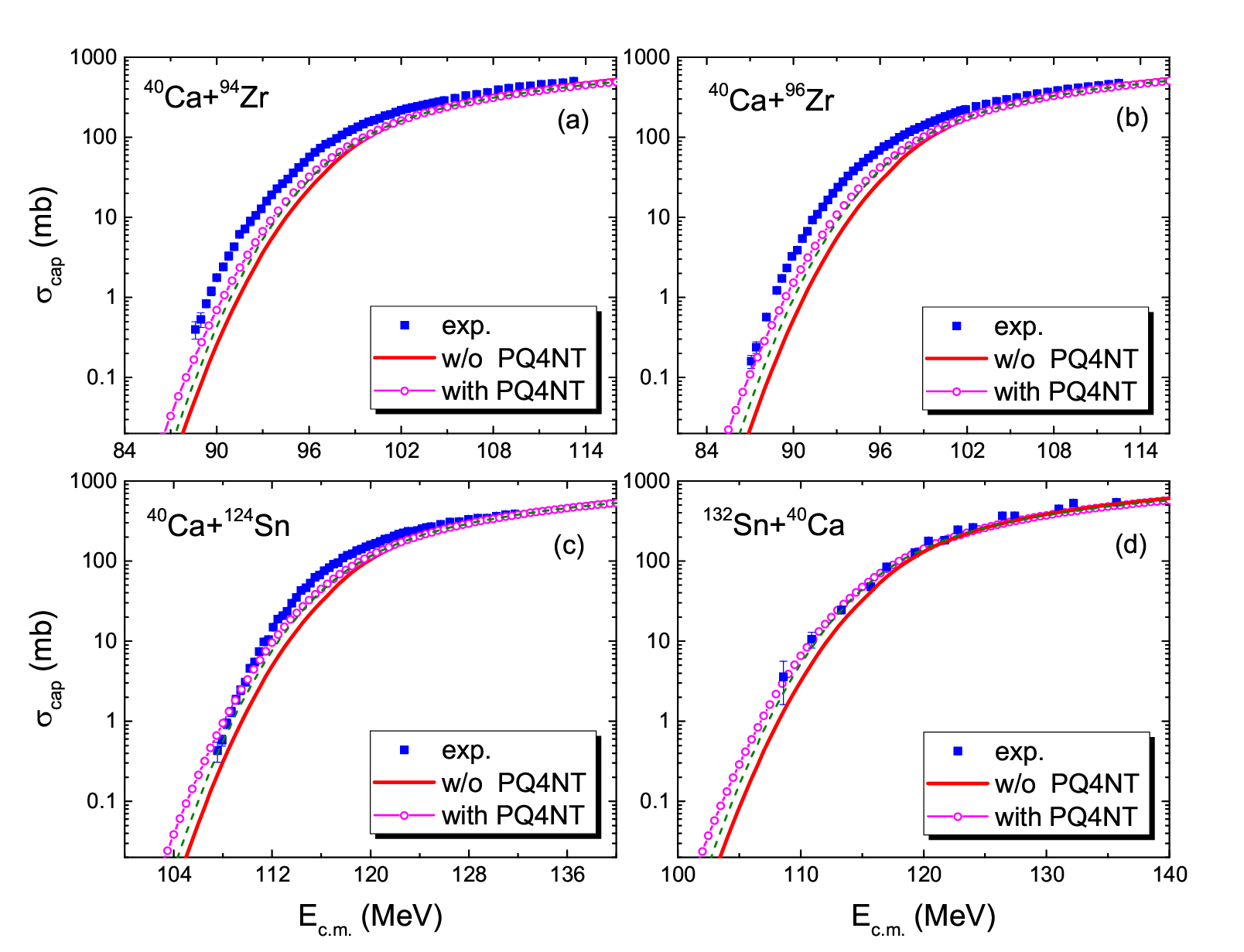}
	\caption{ Capture excitation functions for reactions  $^{40}$Ca + $^{94}$Zr \cite{Sef07},  $^{40}$Ca + $^{96}$Zr \cite{Tim98},  $^{40}$Ca + $^{124}$Sn \cite{Scar00}, and  $^{132}$Sn + $^{40}$Ca \cite{Kola12}. The squares denote the experimental data. The solid curves and the circles denote the results of EBD2 and those with the PQ4NT effect being considered in the calculations, respectively. The dashed curves denote the results considering the  $\Delta_{\rm tr}$ term in Eq.(2) but neglecting the $\Delta_{\rm tr}$ term in Eq.(6).  }
\end{figure}

\begin{figure}
	\setlength{\abovecaptionskip}{ -0.5 cm}
	\includegraphics[angle=0,width=0.8\textwidth]{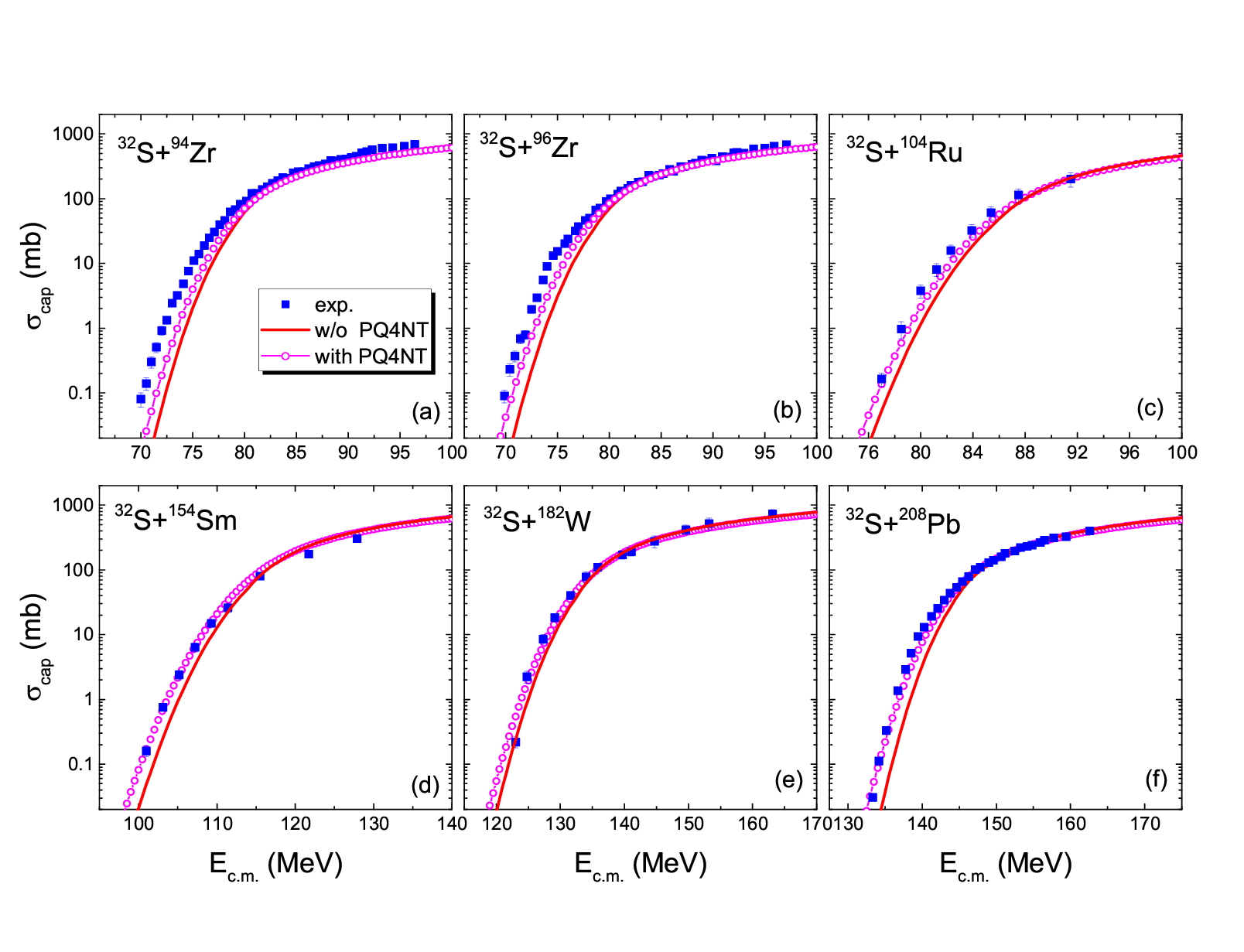}
	\caption{ The same as Fig. 3, but for reactions  $^{32}$S+$^{94}$Zr \cite{Jia14a},  $^{32}$S+$^{96}$Zr \cite{Zhang10},  $^{32}$S+$^{104}$Ru \cite{Pengo83}, $^{32}$S+$^{154}$Sm \cite{Gom94}, $^{32}$S+$^{182}$W \cite{Mis00}, and  $^{32}$S+$^{208}$Pb \cite{Das04}.}
\end{figure}

In Fig. 3, we show the capture excitation functions for four fusion reactions  $^{40}$Ca + $^{94}$Zr,  $^{40}$Ca + $^{96}$Zr,  $^{40}$Ca + $^{124}$Sn, and  $^{132}$Sn + $^{40}$Ca with $Q_{4n}>0$. The circles and the solid curves denote the results with and without the PQ4NT effects being considered, respectively. One sees that the experimental data can be better reproduced with the PQ4NT effects being considered in EBD2 calculations, i.e., the $\Delta_{\rm tr}$ terms are involved. The values of $Q_{4n}$ for these reactions are listed in Table I. We also note that for the reactions with $Q_{4n}>0$, the corresponding $Q$-values for two-neutron transfer are also positive. For $^{40}$Ca + $^{94}$Zr and $^{40}$Ca + $^{96}$Zr, the values of $Q_{4n}$ are 8.13 MeV and 9.64 MeV, respectively, which implies the subsequent neutron-pair transfers are energetically favorable and the enhancements are clearly observed. Considering the PQ4NT effects in the calculations, the results are evidently improved, although the under-prediction of the experimental data still exists. To see the contribution of the $\Delta_{\rm tr}$ term in $W$, we also present the results (the dashed curves) neglecting the $\Delta_{\rm tr}$ term of $W$ in the calculations. We note that the sub-barrier capture cross sections can be further enhanced with the $\Delta_{\rm tr}$ term of $W$ being considered, and the average deviation between the predicted results and 113 datasets of measured cross sections \cite{EBD2} can be further reduced by $6\%$. Over all reactions under consideration in this work, there are only six reactions $^{40}$Ca+$^{192}$Os, $^{40}$Ca+$^{208}$Pb, $^{40}$Ca+$^{238}$U, $^{132}$Sn+$^{40}$Ca, $^{132}$Sn+$^{58}$Ni and $^{134}$Te+$^{40}$Ca with $\Delta_{\rm tr}>1 $ MeV, and the truncation at 1 MeV has minimal impact on the overall fit.

\begin{figure}
	\setlength{\abovecaptionskip}{ -3.5 cm}
	\includegraphics[angle=0,width=0.9\textwidth]{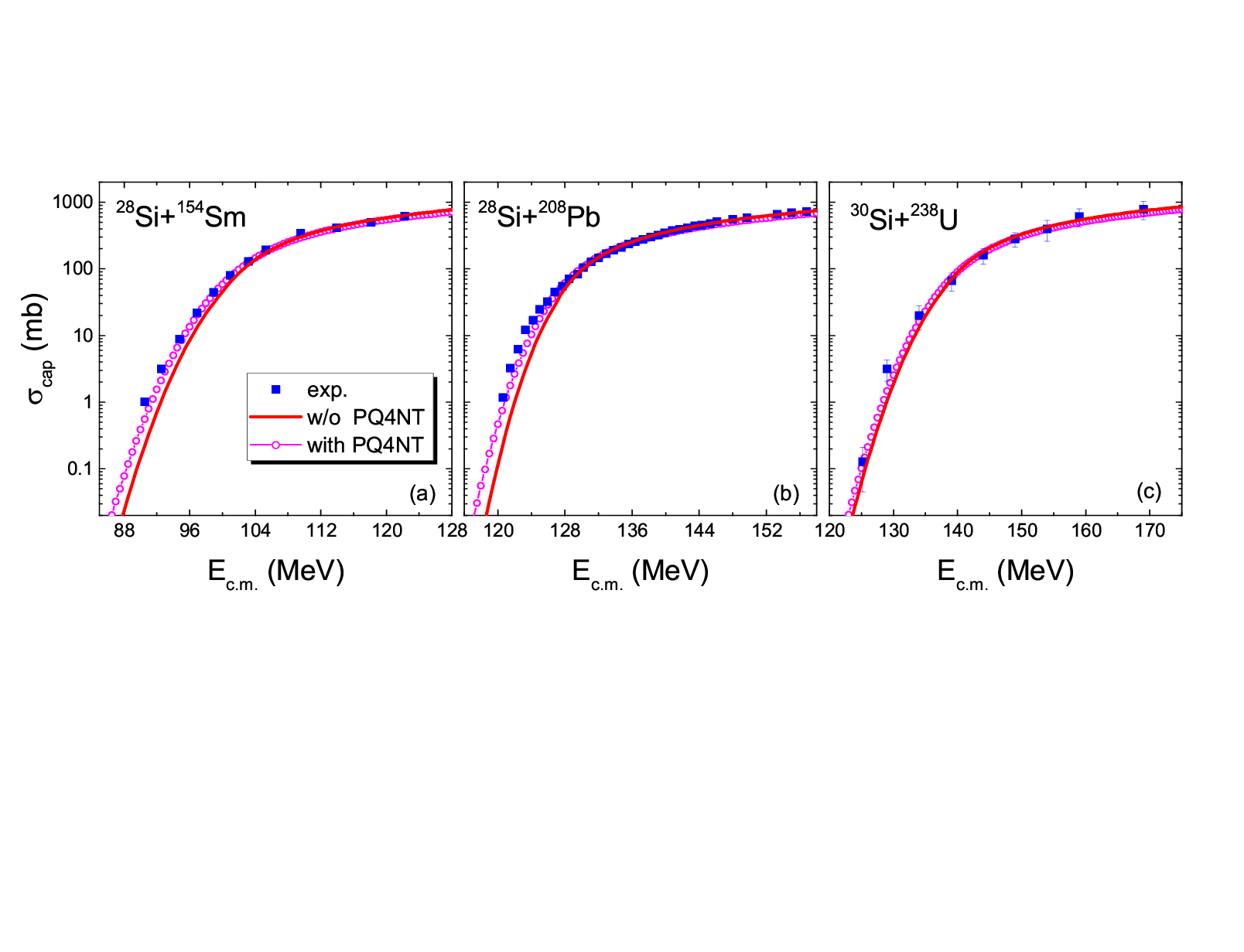}
	\caption{ The same as Fig. 3, but for reactions  $^{28}$Si+$^{154}$Sm \cite{Gil90},  $^{28}$Si+$^{208}$Pb \cite{Hind95} and  $^{30}$Si+$^{238}$U \cite{Nish10}.  }
\end{figure}

In Fig. 4 and Fig. 5, we show the corresponding comparisons for more fusion reactions with $Q_{4n}>0$. One can see that the results are all improved when the PQ4NT effects are involved in the calculations, which indicates that the PQ4NT effects play an important role for the sub-barrier enhancement of the capture cross sections.  
By considering the PQ4NT effects in the calculations for reactions with $Q_{4n}>0$, the average deviation between the predicted results and 113 datasets of measured cross sections  \cite{EBD2} is significantly reduced by $20\%$ (from 0.118 to 0.094). Considering that the values of $Q_{4n}$ are $8.17$ MeV and $14.30$ MeV for $^{36}$Ar+$^{208}$Pb and $^{44}$Ti+$^{208}$Pb (see Table I), respectively, the sub-barrier fusion enhancement due to positive Q-value four-neutron transfer could be evidently observed in these two unmeasured reactions.

\begin{center}
	\textbf{ B. Uncertainties of Model Predictions }\\
\end{center}

In this sub-section, we systematically investigate the uncertainty of the predicted capture cross sections with EBD2.2. Here, a total of 426 datesets of measured cross sections \cite{EBD}, including the systems collected in Refs. \cite{Wangbing,Itkis22,Chen23} except those induced by nuclei lighter than $^{12}$C are used. In Fig. 6(a), we show the deviations between the predicted capture cross sections $\sigma_{\rm th}$ and the experimental data $\sigma_{\rm exp}$ (in logarithmic scale) for 426 datesets of measured cross sections. One can see that the deviations at sub-barrier energies are relatively large, and with the increase of the incident energy the deviations systematically decrease. To investigate the energy-dependence of the uncertainties of model predictions, we calculate the corresponding root-mean-square deviations (RMSD) at different values of $E_{\rm c.m.}/V_{\rm B }$. The squares in Fig. 6(b) shows the calculated RMSD as a function of $E_{\rm c.m.}/V_{\rm B }$. As seen in Fig. 6, the RMSD are relatively larger at sub-barrier energies comparing with the above-barrier cases. At incident energies $E_{\rm c.m.}> 0.9 V_{\rm B }$, the deviation systematically decreases with  $E_{\rm c.m.}$ and approaches to a value of $\sim 0.085$. 

\begin{figure}
	\setlength{\abovecaptionskip}{0 cm}
	\includegraphics[angle=0,width=0.75\textwidth]{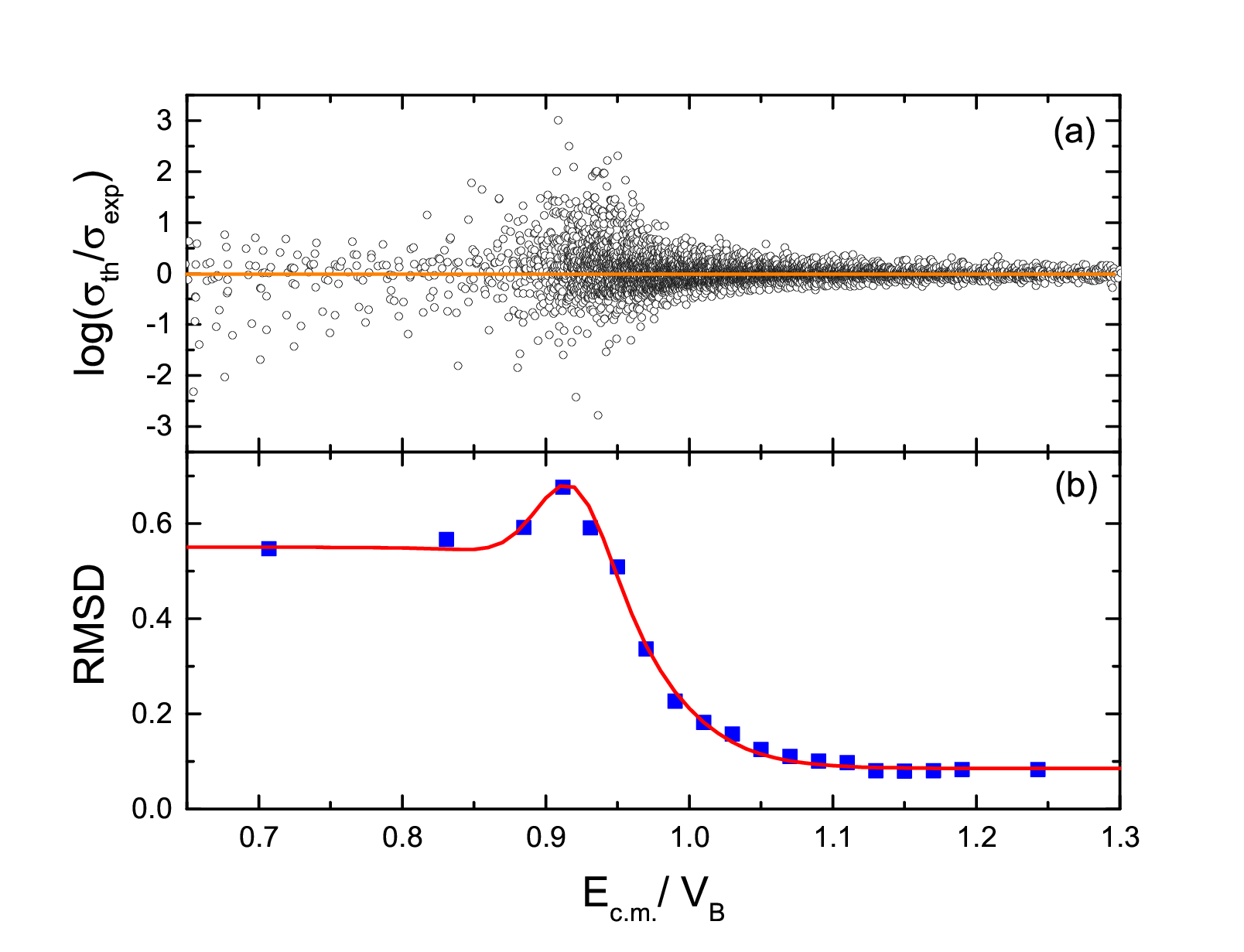}
	\caption{ (a) Deviations between predicted capture cross sections with EBD2.2 and the experimental data as a function of the ratio $E_{\rm c.m.}/V_{\rm B }$. (b) Root-mean-square deviations with respect to the measured capture cross sections (in log) as a function of $E_{\rm c.m.}/V_{\rm B }$. The squares denote the calculated results with the data in (a). The solid curve denotes the results by fitting the squares.  }
\end{figure}

By fitting the calculated rms deviations with a Gaussian function plus a Fermi function, we obtain an energy-dependent RMSD for EBD2.2 [see the solid curve in Fig. 6(b)], 
\begin{eqnarray}
{\rm RMSD}=0.085  + 0.20 \exp \left [  - (\frac{E_{\rm c.m.}/V_{\rm B }-0.92}{0.035})^2 \right ] + 0.465 \left [1+\exp(\frac{E_{\rm c.m.}/V_{\rm B }-0.97}{0.03}) \right ]^{-1}.
\end{eqnarray}  
Equation (8) provides an energy-dependent root-mean-square deviation (RMSD) derived from the extensive datasets of measured fusion reactions. This empirical RMSD can be applied to estimate the uncertainty of predictions for unmeasured systems. Fig. 7 shows the predicted capture cross sections for six fusion reactions with deformed nuclei, in which several reactions are not included in the fitting datasets. For energies well above the barrier, the RMSD converges to approximately 0.085, corresponding to a relative uncertainty of about $21.6\%$ (i.e., $\sigma_{\rm th} \times 10^{0.085} \approx 1.216 \sigma_{\rm th} $). This quantitative uncertainty assessment is particularly valuable for predicting evaporation residue cross sections in the synthesis of superheavy nuclei, where accurate extrapolations of capture cross sections are crucial \cite{JJLi25}. To test the predictive power of EBD2.2, the measured capture excitation functions of three reactions $^{12}$C+$^{24}$Mg \cite{Mon22,Gary82}, $^{12}$C+$^{184}$W \cite{San22} and $^{12}$C+$^{248}$Cm \cite{Ban20} which are not included in the 426 datasets mentioned previously, are also presented in Fig. 7 for comparison. The shadows denote the predictions of EBD2.2 with default values for the parameters. One sees that these "new" data can be reproduced reasonably well.

\begin{figure}
	\setlength{\abovecaptionskip}{0 cm}
	\includegraphics[angle=0,width=0.95\textwidth]{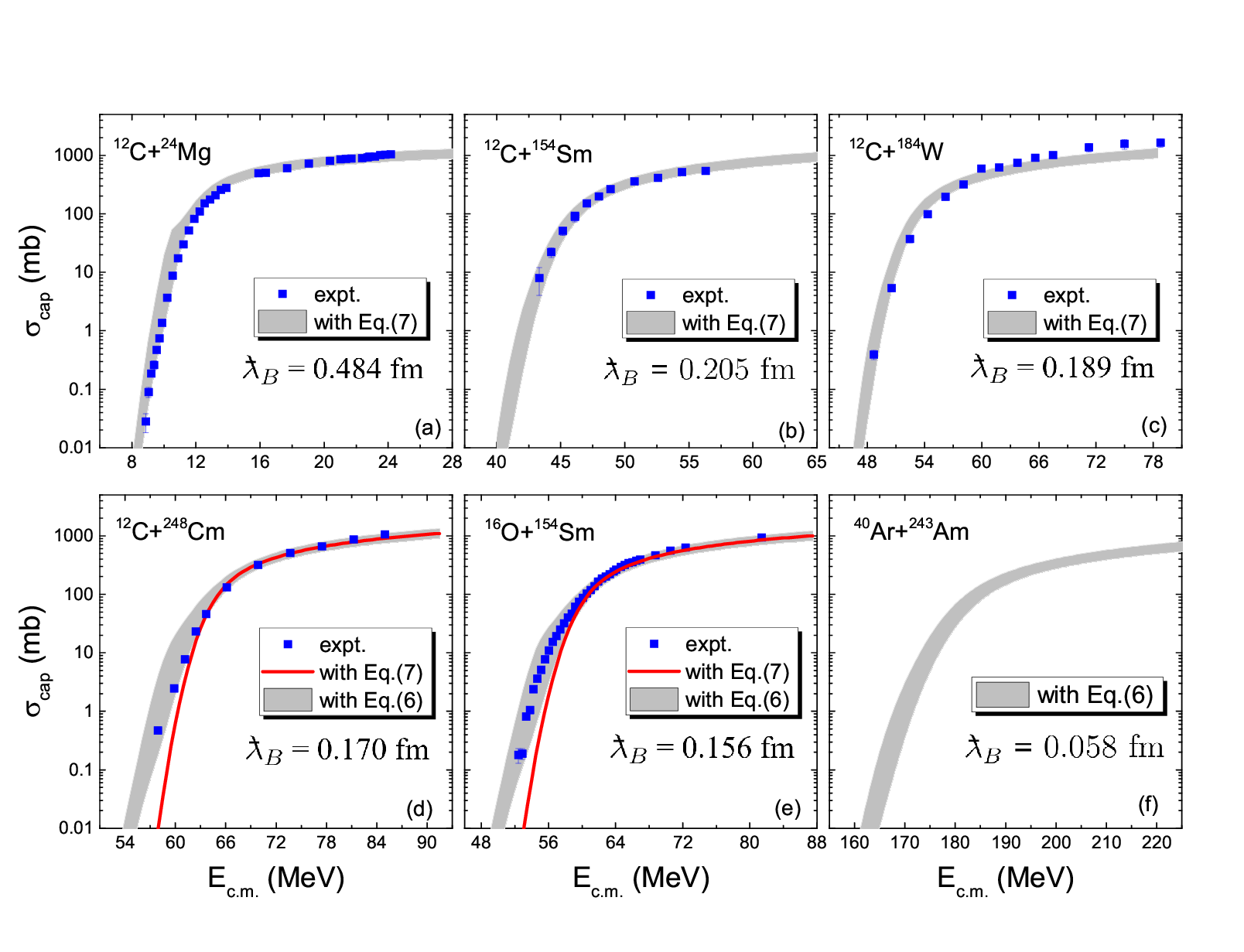}
	\caption{ Fusion excitation functions for six fusion reactions with deformed nuclei. The shadow denotes the uncertainty of EBD2.2 predictions. The squares denote the experimental data taken from Refs. \cite{Mon22,Gary82,Gil85,Lei95,San22,Ban20}. }
\end{figure}

To investigate the influence of the de Broglie wavelength on the capture cross sections, we also present in Fig. 7 the reduced de Broglie wavelength $\middlebar{\lambda}_B$ of the colliding nuclei. For $^{12}$C+$^{24}$Mg, $^{12}$C+$^{154}$Sm and $^{12}$C+$^{184}$W, the values of $\middlebar{\lambda}_B$ are larger than 0.18 fm. One sees that the experimental data for these three reactions can be well reproduced by using Eq.(7) in the calculations, i.e., neglecting the deformation effects. It is known that as the de Broglie wavelength increases for light systems, the wave function of the incident nucleus spreads out, such that it is no longer sensitive to specific orientations of the deformed target nucleus, but instead interacts with an average potential field. As a result, the enhancement of the capture cross sections at sub-barrier energies due to deformation weakens, and its behavior gradually approaches the results calculated using spherical shapes. Whereas, for heavier systems such as $^{12}$C+$^{248}$Cm (with $\middlebar{\lambda}_B=0.170$ fm) and $^{16}$O+$^{154}$Sm (with $\middlebar{\lambda}_B=0.156$ fm), we note that the measured cross sections are under-predicted if neglecting the deformation effects, which indicates that the light nucleus with de Broglie wavelength of about $\middlebar{\lambda}_B<0.18$ fm can "sense" the anisotropic effects due to the orientations of the deformed target nucleus and the deformation effects therefore cannot be neglected. In Fig. 7(f), the capture excitation functions for $^{40}$Ar+$^{243}$Am are simultaneously predicted, in which $\middlebar{\lambda}_B$ is very small and the deformation effects play a role to the sub-barrier capture cross sections.

\begin{center}
	\textbf{C. Physical Interpretation of PQ4NT Effects}
\end{center}

The sub-barrier fusion enhancement due to neutron transfer was previously investigated based on QCC+ENR approach, i.e., quantum coupled-channels calculations together with the semiclassical relations for the neutron transfer probabilities \cite{Karpov15}. It is found that the neutron rearrangement in four PQ4NT systems $^{32}$S+$^{94,96}$Zr  and $^{40}$Ca+$^{94,96}$Zr leads to broadening of the barrier distribution functions. The low-energy tails (see Fig. 2 and Fig. 3 in \cite{Karpov15}) due to the neutron transfer significantly enhance the sub-barrier capture cross sections. In EBD2.2, the PQ4NT effects make the average barrier height lower and the width of the barrier distribution wider when $Q_{4n} > 0$, which is in good agreement with the conclusions from QCC+ENR. Zagrebaev also noted that the neutron transfer effects in $^{40}$Ca+$^{48}$Ca make the most probable barrier height lower and the width of the barrier distribution wider \cite{Zag03}. In addition, the neutron transfer is clearly observed from the improved quantum molecular dynamics simulations for $^{132}$Sn+$^{40}$Ca, and the ratio of neutron-density to proton-density at neck is higher than the $N/Z$ of the compound nucleus by a factor of two when the reaction partners begin to contact each other \cite{ImQMD2014}.

 \begin{table}    	
	
	\caption{ Capture barrier heights for reactions with $^{40,48}$Ca bombarding on doubly-magic nuclei.   }
	\begin{tabular}{ccccc}
		\hline\hline
		
		~~~Reaction~~~  & ~~~$V_{\rm B }^{\rm TDHF}$ (MeV)~~~ &~~~$V_{\rm FD}$ (MeV)~~~  &  ~~~ $V_{\rm B }^{\rm TDHF}/ \, V_{\rm FD}$ ~~~  & ~~~$Q_{4n}$ (MeV) ~~~    \\
		\hline
		$^{40}$Ca+$^{48}$Ca  &  51.98   &  54.81   &  0.948   &  $+3.87$     \\
		$^{40}$Ca+$^{96}$Zr  &  95.07   &  101.53  &  0.936   &  $+9.64$     \\
		$^{40}$Ca+$^{132}$Sn &  114.38  &  120.28  &  0.951   &  $+13.44$   \\
		$^{40}$Ca+$^{208}$Pb &  176.91  &  185.51  &  0.954   &  $+9.98$     \\

		$^{48}$Ca+$^{48}$Ca  &  51.12   &  53.52   &  0.955   &  $-12.72$     \\
		$^{48}$Ca+$^{96}$Zr  &  93.51   &  99.32   &  0.942   &  $-6.95$     \\
		$^{48}$Ca+$^{132}$Sn &  112.66  &  118.32  &  0.952   &  $-3.14$     \\
		$^{48}$Ca+$^{208}$Pb &  174.78  &  182.32  &  0.959   &  $-6.60$     \\
		
		\hline\hline
	\end{tabular}
\end{table}

To further understand the PQ4NT effects, we investigate the capture barrier height by using the microscopic TDHF theory. Based on the TDHF code Sky3D \cite{Maru14}, the capture barrier heights $V_{\rm B }^{\rm TDHF} $\cite{Yao24} for four groups of fusion reactions $^{40,48}$Ca+$^{48}$Ca, $^{40,48}$Ca+$^{132}$Sn, $^{40,48}$Ca+$^{208}$Pb and $^{40,48}$Ca+$^{96}$Zr  are calculated by using the Skyrme energy density functional with the parameter set SLy6 \cite{SLy6}.  
In Table II, we show the calculated results. Here, $V_{\rm FD}$ denotes the corresponding frozen-density barrier height obtained by using the same Skyrme energy density functional together with the extended Thomas-Fermi approach \cite{liumin}. The values of $Q_{4n}$ are also presented for comparison. 
While TDHF calculations encompass a variety of microscopic effects (including proton transfer, nucleon exchange, and collective excitations), the systematic reduction of $V_{\rm B }^{\rm TDHF}/\, V_{\rm FD}$ for reactions with $Q_{4n}>0$ compared to those with $Q_{4n}<0$ suggests that the four-neutron transfer plays a significant role in lowering the dynamical barrier when it is energetically favorable. This observation aligns with the PQ4NT hypothesis that sustained neutron transfer facilitates barrier reduction. From Table II, we note that the values of $Q_{4n}$ are all positive for $^{40}$Ca-induced reactions and  $Q_{4n}$ are all negative for $^{48}$Ca-induced reactions, and the ratio $V_{\rm B }^{\rm TDHF}/\, V_{\rm FD}$ in $^{40}$Ca-induced reactions are systematically smaller than those in $^{48}$Ca-induced reactions. This indicates that the PQ4NT effect contributes significantly to the additional reduction of the dynamical barrier in these reactions.  

\begin{figure}
	\setlength{\abovecaptionskip}{-4 cm}
	\includegraphics[angle=0,width=0.85\textwidth]{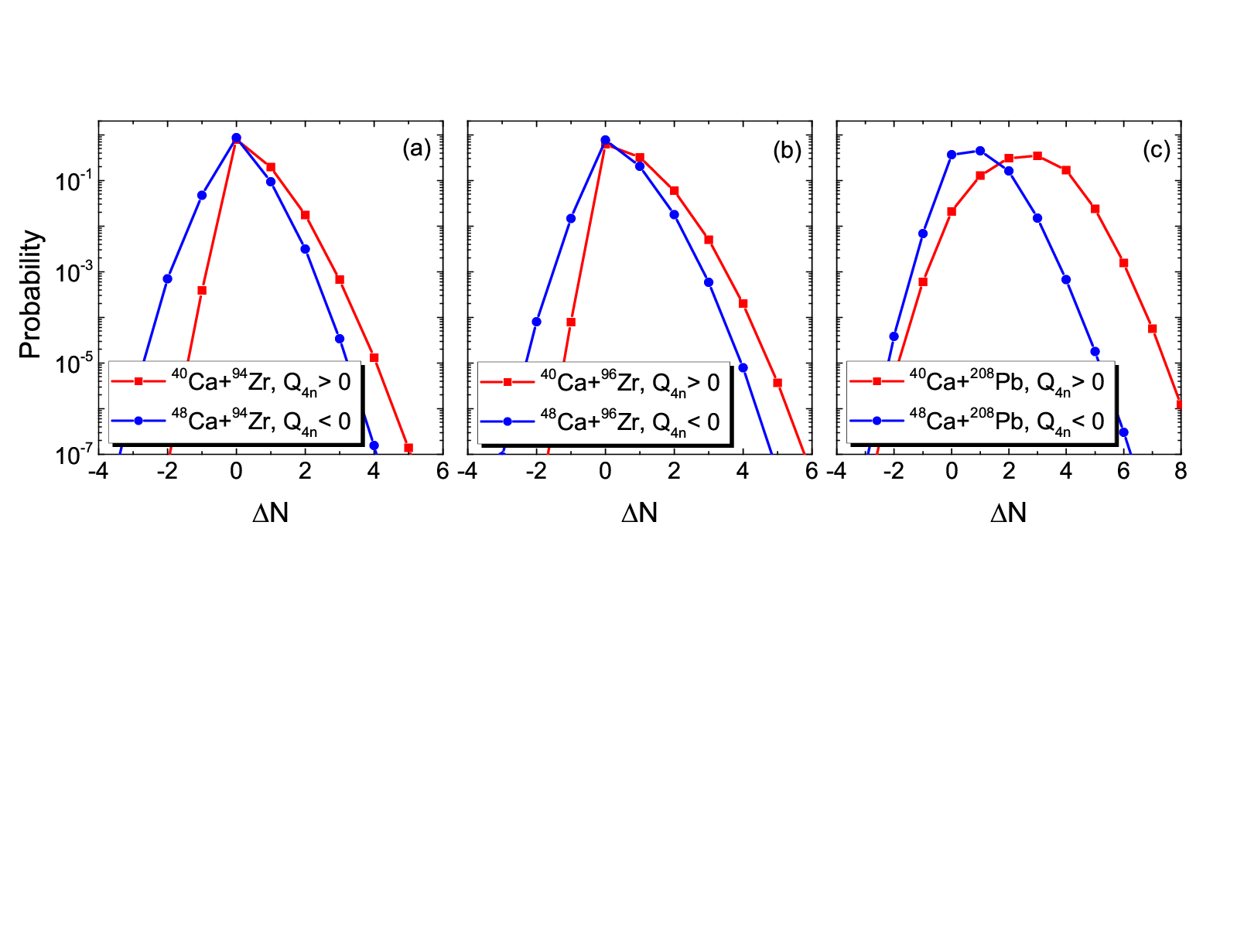}
	\caption{  Neutron transfer probability as a function of the number of neutrons transferred from the target nucleus to the projectile nucleus in back-angle quasi-elastic scattering of $^{40,48}$Ca+$^{94,96}$Zr and $^{40,48}$Ca+$^{208}$Pb. The squares and the circles denote the results for reactions with $Q_{4n}>0$ and those with $Q_{4n}<0$, respectively. }
\end{figure}

It is found that in addition to the fusion barrier distribution directly extracted from the measured fusion excitation function, a similar barrier distribution can also be extracted from quasi-elastic scattering (a sum of elastic, inelastic, and transfer processes) at backward angles \cite{Hag08,Tim95,Tim98}. Because the fusion and the quasi-elastic scattering is related to each other due to the flux conservation, similar information can be obtained from those processes and the similarity between the two representations for the barrier distribution has been shown to hold for some intermediate mass systems \cite{Hag08}. In this work, we also investigate the neutron transfer in back-angle quasi-elastic scattering of $^{40,48}$Ca+$^{94,96}$Zr and $^{40,48}$Ca+$^{208}$Pb by using the TDHF theory together with the particle number projection approach \cite{Guo19,Sim10}. Here, the incident energies in the calculations are set as $E_{\rm c.m.}=96.02$ MeV, 93.55 MeV, 95.32 MeV, 93.13 MeV, 176.00 MeV and 174.00 MeV for $^{40}$Ca+$^{94}$Zr, $^{48}$Ca+$^{94}$Zr, $^{40}$Ca+$^{96}$Zr, $^{48}$Ca+$^{96}$Zr, $^{40}$Ca+$^{208}$Pb and $^{48}$Ca+$^{208}$Pb, respectively, which are slightly lower than the corresponding capture thresholds.  Fig. 8 shows the transfer probability of neutrons transferred from target to projectile ($\Delta N$). Where $\Delta N>0$ indicates neutrons being transferred from the target to the projectile and $\Delta N<0$ indicates neutrons being transferred from the projectile to the target. From Fig. 8, one sees that for systems with $Q_{4n}>0$ (exothermic reactions), such as $^{40}$Ca+$^{94,96}$Zr and $^{40}$Ca+$^{208}$Pb, the probability at $\Delta N=4$  is much higher than that for the corresponding systems with $Q_{4n}<0$ (endothermic reactions). For example, the probability of $^{40}$Ca+$^{94}$Zr at $\Delta N=4$ is approximately 10$^{-5}$, while that of $^{48}$Ca+$^{94}$Zr is only 10$^{-7}$. A higher probability for 4n transfer observed in the back-angle quasi-elastic scattering provides independent and direct microscopic evidence for the propensity of this multi-neutron process to occur as the nuclei approach the barrier. This propensity is fundamentally linked to the mechanism of neutron-rich neck formation, which subsequently lowers the dynamical fusion barrier. 

The PQ4NT effect influences the fusion barrier in two interrelated ways: lowering the average barrier height and broadening the barrier distribution. During the approach of the colliding nuclei, the transfer of four neutrons from the neutron-rich partner to the neutron-deficient one is energetically favorable when $Q_{4n}>0$. This sustained neutron transfer leads to the formation of a neutron-rich neck between the nuclei, which reduces the Coulomb repulsion and thus the effective barrier height. Concurrently, the transfer process introduces additional degrees of freedom and dynamical fluctuations, which manifest as a wider barrier distribution. In the EBD2.2 parametrization, these effects are captured by the correction terms $\Delta_{\rm tr}$ in both the average barrier height $V_B$  and the width $W$. The term $\Delta_{\rm tr}=Q_{4n}|I_1-I_2|/2$ quantifies the energy gain from the four-neutron transfer weighted by the isospin asymmetry difference, reflecting the propensity for neutron flow between the reaction partners. The inclusion of $\Delta_{\rm tr}$ in $W$ further accounts for the smearing of the barrier due to the neutron transfer dynamics, consistent with the observed enhancement of sub-barrier capture cross sections.

\begin{center}
	\textbf{IV. SUMMARY}
\end{center}

 In this work, the influence of positive $Q$-value four-neutron transfer (PQ4NT) on sub-barrier capture cross sections has been systematically examined within the empirical barrier distribution (EBD2) framework. It is shown that for reactions with $Q_{4n}>0$, such as $^{28}$Si, $^{32}$S+$^{154}$Sm and $^{40}$Ca+$^{94,96}$Zr, sustained neutron-pair transfer leads to a reduction in the fusion barrier height and a concomitant enhancement of capture cross sections at sub-barrier energies. In contrast, systems with positive $Q_{2n}$ but negative $Q_{4n}$ — exemplified by $^{18}$O+$^{58}$Ni and  $^{18}$O+$^{116}$Sn — exhibit no discernible enhancement, as neutron transfer ceases after the initial pairs. The original EBD2 model accurately reproduces experimental data for reactions with $Q_{4n}<0$, including $^{16}$O+$^{154}$Sm and $^{48}$Ca+$^{154}$Sm. However, it systematically underestimates sub-barrier cross sections for reactions with $Q_{4n}>0$. By incorporating the PQ4NT effect through modifications to both the barrier height and distribution width, the predicted capture excitation functions for all 13 studied systems show significantly improved agreement with experimental data. Furthermore, TDHF calculations indicate greater dynamic barrier reduction in $^{40}$Ca-induced reactions (with $Q_{4n}>0$) compared to $^{48}$Ca-induced ones, and much larger neutron pickup probabilities are observed in the quasi-elastic scattering reactions induced by $^{40}$Ca.  Predictions for as-yet unmeasured systems such as $^{36}$Ar+$^{208}$Pb   and $^{44}$Ti+$^{208}$Pb also suggest observable PQ4NT-induced enhancement. These results underscore the critical role of multi-neutron transfer continuity in sub-barrier fusion dynamics. The proposed model extensions yield a $20\%$ reduction in cross-section deviation, demonstrating notable predictive improvement. We have also quantified the predictive uncertainty of the model by deriving an energy-dependent RMSD. The uncertainties are within half an order of magnitude at sub-barrier energies and approximately $21.6\%$ at energies well above the barrier.

\begin{center}
	\textbf{ACKNOWLEDGEMENTS}
\end{center}
This work was supported by  Guangxi "Bagui Scholar" Teams for Innovation and Research Project, National Natural Science Foundation of
China (Nos. 12265006, U1867212), and  National Key R$\&$D Program of China (Grants No. 2023YFA1606402). Online calculations with EBD2.2 are available on http://www.imqmd.com/fusion/EBD2B.html

\end{document}